\documentclass[conference]{IEEEtran}
\IEEEoverridecommandlockouts
\usepackage{cite}
\usepackage{amsmath,amssymb}
\usepackage{graphicx}
\usepackage{booktabs}
\usepackage{url}
\usepackage[caption=false,font=footnotesize]{subfig}
\graphicspath{{figs/}}

\begin{document}

\title{A Quad Beam Variable Height Antenna System for Reliable Underground Mine Tunnel Communications
\thanks{The work was supported by the National Science Foundation (NSF) under Award no. CCF-2621730. (Corresponding author: Sihua Shao.)}}

\author{\IEEEauthorblockN{Sihua Shao, Charles Fent, Matthew Dodd, and Atef Elsherbeni}
\IEEEauthorblockA{Department of Electrical Engineering, Colorado School of Mines, Golden, CO 80401, USA\\
Emails: sihua.shao@mines.edu, charles\_fent@mines.edu, mdodd@mines.edu, aelsherb@mines.edu}}

\maketitle

\begin{abstract}
Mine accidents can damage the wired backbone supporting underground wireless networks, forcing rescue teams to deploy temporary mesh nodes to maintain communication with the surface. Although higher carrier frequencies can support bandwidth-intensive applications such as live video, multimode propagation within mine tunnels produces severe, spatially localized fading. Movable antennas can mitigate these fades by adjusting antenna position; however, moving all antennas on a multi-link relay node simultaneously may improve one link at the expense of another. This paper introduces QuadBeam, a compact mesh node antenna system comprising four independently positionable 2.4~GHz patch antennas mounted on the side faces of a central hub, together with a 900~MHz monopole for backup voice communication. A preliminary prototype incorporates one motor-driven patch mount and three manually adjustable mounts. Bench measurements show interelement coupling below $-46$~dB between patches on adjacent faces. Measurements at five carrier frequencies from 2.40~GHz to 2.48~GHz in the Edgar Experimental Mine reveal 19~dB to 26~dB received-power variations across transmitter and receiver separations of 8.3~m to 18.3~m. Adjusting the receiver height over approximately 20~cm yields received-power improvements of up to 22~dB. Repositioning neighboring patches and removing or reinstalling the monopole produce no noticeable change in the active link. These results demonstrate the potential of independently positioned antennas to alleviate spatial fading in underground mesh communications.
\end{abstract}


\section{Introduction}

\begin{figure}[t]
\centering
\includegraphics[width=\columnwidth]{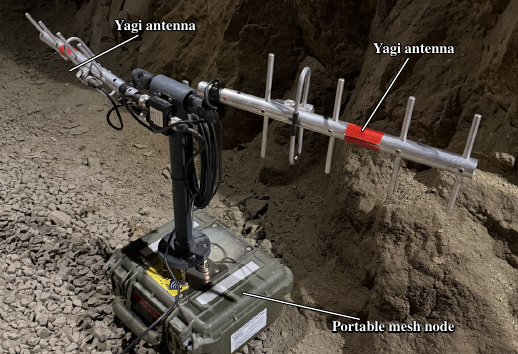}
\caption{A portable mesh node for mine rescue (IWT SENTINEL) in a mine tunnel. Its two Yagi antennas point in opposite directions along the tunnel.}
\label{fig:sentinel}
\end{figure}

Explosions, fires, and roof falls in underground mines can disrupt existing communication infrastructure~\cite{duane2023,duane2026}, including wireless systems that depend on a wired, typically fiber-optic, backbone. Rescue teams must then establish temporary connectivity using battery-powered mesh nodes deployed progressively along the tunnels. These nodes relay voice, messaging, and personnel-tracking information between rescue teams and a surface command center~\cite{iwt_minerescue}. Directional antennas are commonly employed to extend communication along the tunnel axis, as illustrated by the portable node in Fig.~\ref{fig:sentinel}.

While voice and tracking require relatively modest bandwidth, emerging applications including live video and robotic lidar mapping may require data rates of tens of megabits per second~\cite{duane2023}. Operation at 2.4~GHz and above offers access to wider channels and can reduce longitudinal attenuation under some tunnel-propagation conditions~\cite{sun2010,zhou2015}. However, a mine tunnel behaves as a large, lossy multimode waveguide. Interference among propagating modes produces severe spatial fading, particularly in the near zone, and the received field depends strongly on the positions of both antennas within the tunnel cross section~\cite{sun2010,zhou2015,tardioli2019,zhou2015aps,rizzo2019}. Consequently, a node deployed at an unfavorable location may experience a persistent deep fade.

Movable antennas (MAs)~\cite{zhu2024twc,zhu2026tutorial} and related fluid-antenna systems~\cite{wong2021} address spatial fading by repositioning an antenna within a limited region to exploit local variations in channel strength without adding radio-frequency chains. Prior experiments have demonstrated large received-power variations in indoor environments~\cite{dong2026}, while an underground coal-mine study considered movable antennas in conjunction with a reconfigurable intelligent surface through simulation~\cite{xia2026}. To the authors' knowledge, experimental validation of movable-antenna operation within an underground mine has not previously been reported.

A multi-link mesh relay introduces a challenge that does not arise in a single-link configuration. A relay must maintain connections to multiple neighboring nodes, potentially in different directions. Because each link experiences a distinct spatial fading profile, a common displacement of all antennas may improve one connection while degrading another. QuadBeam addresses this limitation by providing independently adjustable antennas: four 2.4~GHz patches positioned on the four side faces of a compact hub, supplemented by a 900~MHz monopole intended for backup voice service.

This paper presents the first QuadBeam prototype and preliminary experimental results. One patch is driven by a linear actuator, whereas the remaining mounts are manually adjustable. For controlled mine measurements, the complete receiver hub was raised or lowered on a nonmetallic cart. The contributions are (i) demonstration of low coupling between neighboring antennas, (ii) characterization of frequency-selective spatial fading over short distances in an underground mine drift, and (iii) evaluation of the received-power recovery achievable through modest vertical repositioning. Section~\ref{sec:prototype} describes the antenna design and bench characterization; Section~\ref{sec:edgar} presents measurements in the Edgar Experimental Mine; and Section~\ref{sec:conclusion} summarizes the findings and future work.

\section{QuadBeam Antenna Prototype}
\label{sec:prototype}

\begin{figure*}[t]
\centering
\includegraphics[width=\textwidth]{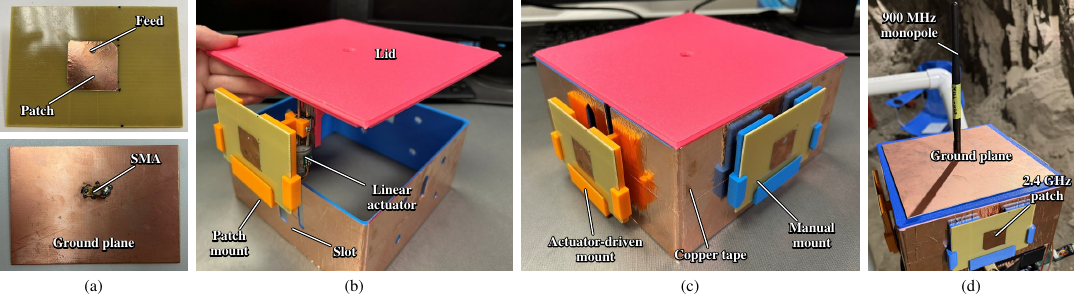}
\caption{QuadBeam prototype. (a) Patch antenna: copper-tape patch on FR4 (top) and ground plane with the SMA connector (bottom). (b) Linear actuator under the lid of the hub and the patch mount that it drives through a slot in the wall. (c) Hub with an actuator-driven mount and a manual mount on adjacent faces. (d) Complete prototype in the mine, with the 900~MHz monopole on its ground plane.}
\label{fig:prototype}
\end{figure*}

The prototype consists of a compact hub, four nominally identical 2.4~GHz patch antennas, and a 900~MHz monopole (Fig.~\ref{fig:prototype}). Five patch elements were fabricated so that one could be used separately as the transmitting antenna during the measurements described in Section~\ref{sec:edgar}.

\subsection{Antenna Design and Fabrication}
\label{sec:fabrication}

\subsubsection{Patch antennas}
Each antenna is a single-feed square microstrip patch with two truncated corners to produce right-hand circular polarization (RHCP), as shown in Fig.~\ref{fig:prototype}(a). Circular polarization reduces sensitivity to uncontrolled node orientation: a circularly polarized antenna receives a linearly polarized field with an ideal 3~dB polarization mismatch, whereas orthogonally polarized linear antennas can experience much larger penalties~\cite{zhou2015}. The design was optimized in Ansys HFSS following~\cite{habib2015}, beginning with a patch dimension of approximately half the substrate wavelength, followed by feed-location and corner-truncation sweeps to improve impedance matching and axial ratio.

The final patch has a side length of 28.6~mm, 2~mm corner truncations, and a feed offset of 5.89~mm in one direction and 0.98~mm in the other. It is fabricated on a $100 \times 70 \times 1.5$~mm$^3$ single-sided copper-clad FR4 board with nominal relative permittivity 4.4; the copper-clad surface forms the ground plane. The radiating patch is cut from 0.05~mm copper foil tape with a desktop cutting machine and applied to the opposite substrate face. A 1.6~mm feed hole, cleared ground region, and SMA connector complete the assembly. After initial measurements indicated a resonance shift, the substrate model was refined and the design was iterated.

\subsubsection{Monopole}
A commercial 900~MHz monopole is connected through an SMA adapter to an $18 \times 18$~cm$^2$ copper-clad ground plane mounted above the hub. The ground-plane side length is slightly greater than half a free-space wavelength at 915~MHz.

\subsubsection{Hub and positioning mechanism}
The 3-D-printed polylactic-acid hub has a footprint matching the monopole ground plane and a height of 10~cm. Copper foil on each side wall provides shielding between patch antennas and the electronics housed within the hub. The four patch mounts move vertically along the side faces. Although the intended system uses an actuator for each patch, the present prototype incorporates one miniature stepper-motor linear actuator with 34~mm travel (Fig.~\ref{fig:prototype}(b)); the other three mounts are manually repositioned (Fig.~\ref{fig:prototype}(c)). The distinction between the proposed four-actuator configuration and the current single-actuator prototype is important when interpreting the experimental results.

\subsection{Antenna Characterization}
\label{sec:testing}

\subsubsection{Reflection coefficient}
\begin{figure}[t]
\centering
\includegraphics[width=\columnwidth]{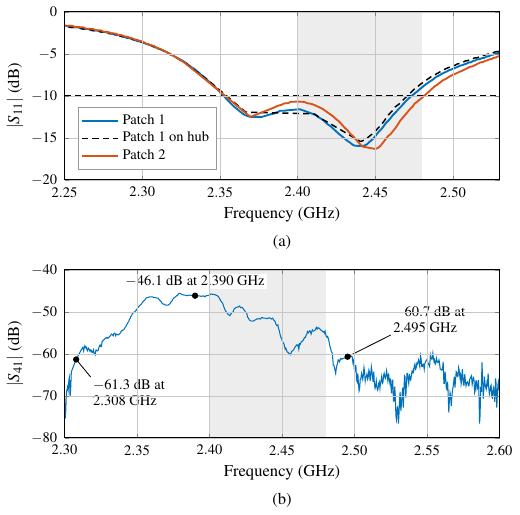}
\caption{Measurements with the VNA. (a) Reflection coefficient of Patches~1 and 2, and of Patch~1 mounted on the hub. (b) Coupling between Patches~1 and 2 on adjacent faces of the hub; the dots are the marker readings of the VNA. The shaded band marks the carriers used in the mine.}
\label{fig:vna}
\end{figure}

Each antenna was characterized using a Copper Mountain Technologies 808/1 vector network analyzer (VNA) following a one-port calibration. Table~\ref{tab:s11} summarizes the measured $-10$~dB impedance bandwidths and the minimum reflection coefficients. The monopole covers 892--933~MHz. Patches~1 and 2, used for the underground transmitter and receiver, cover 2350--2474~MHz and 2350--2482~MHz, respectively (Fig.~\ref{fig:vna}(a)). At the five test frequencies, their measured reflection coefficients range approximately from $-9$~dB to $-16$~dB. Patches~3--5 exhibited greater fabrication-related variation; Patch~5 had the narrowest $-10$~dB band (57~MHz). Installing Patch~1 on the hub changed its reflection coefficient by less than 1~dB.

\begin{table}[t]
\caption{Measured Antenna Impedance Characteristics}
\label{tab:s11}
\centering
\begin{tabular}{@{}lcccc@{}}
\toprule
 & $-10$~dB band & Bandwidth & \multicolumn{2}{c}{Minimum $|S_{11}|$} \\
\cmidrule(l){4-5}
Antenna & (MHz) & (MHz) & (dB) & at (MHz) \\
\midrule
Monopole & 892--933 & 41 & $-28$ & 912 \\
Patch 1\textsuperscript{a} & 2350--2474 & 124 & $-16$ & 2439 \\
Patch 2\textsuperscript{a} & 2350--2482 & 132 & $-16$ & 2446 \\
Patch 3 & 2361--2430 & 69 & $-19$ & 2402 \\
Patch 4 & 2377--2493 & 116 & $-14$ & 2461 \\
Patch 5 & 2423--2480 & 57 & $-12$ & 2456 \\
\bottomrule
\multicolumn{5}{@{}l}{\footnotesize \textsuperscript{a}Transmitting and receiving antennas in Section~\ref{sec:edgar}.} \\
\end{tabular}
\end{table}

\subsubsection{Radiation properties}
\begin{figure}[t]
\centering
\includegraphics[width=\columnwidth]{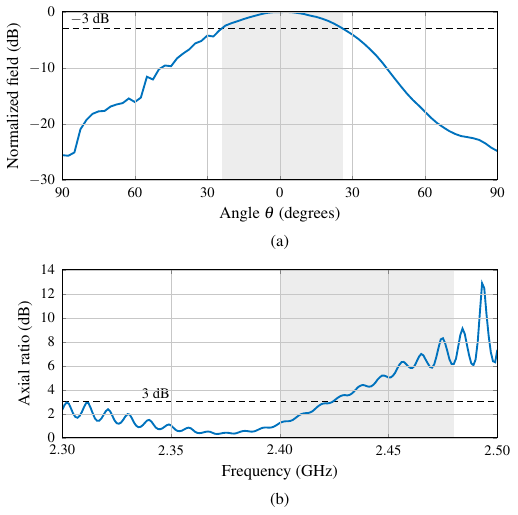}
\caption{Measurements in the anechoic chamber. (a) Normalized E-field pattern at 2.40~GHz versus the angle $\theta$ from boresight; the shaded band is the half-power beamwidth. (b) Axial ratio at boresight; the shaded band marks the carriers used in the mine.}
\label{fig:chamber}
\end{figure}

Anechoic-chamber measurements of a representative patch show a single main lobe with a half-power beamwidth of approximately 50$^\circ$ at 2.40~GHz (Fig.~\ref{fig:chamber}(a)). The boresight axial ratio (Fig.~\ref{fig:chamber}(b)) is below 3~dB from 2.30~GHz to 2.42~GHz, reaching approximately 0.3~dB near 2.37~GHz. At the test frequencies, the axial ratio increases from 1.2~dB at 2.40~GHz to approximately 6~dB at 2.46--2.48~GHz. Thus, the antenna maintains near-circular polarization at the lower frequencies but becomes elliptically polarized toward the upper end of the measurement band.

\subsubsection{Interelement coupling}
Patches~1 and 2 were installed on adjacent hub faces and connected to VNA ports 1 and 4. The measured transmission coefficient $|S_{41}|$ was approximately $-46.1$~dB at 2.39~GHz and fell below $-50$~dB above 2.43~GHz (Fig.~\ref{fig:vna}(b)), consistent with less than 0.003\% coupled power at the highest measured coupling. A 50~$\Omega$ termination measurement gave readings between $-75$ and $-100$~dB, confirming that the observed signal was above the instrument noise floor. These bench measurements were obtained outside an anechoic chamber and without the monopole installed. Coupling between opposite faces and between the patches and monopole was therefore assessed only indirectly through the field check reported in Section~\ref{sec:results}.

\section{Edgar Experimental Mine Measurements}
\label{sec:edgar}

\begin{figure*}[t]
\centering
\includegraphics[width=\textwidth]{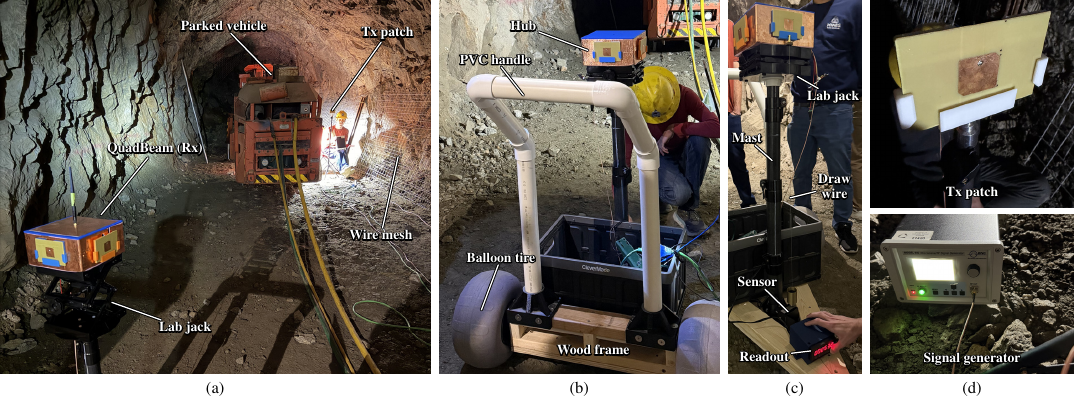}
\caption{Measurement platform in the Edgar Experimental Mine. (a) Line-of-sight link in the drift, with the QuadBeam receiver (Rx) in the foreground and the transmit (Tx) patch beside a parked vehicle. (b) Nonmetallic mobility cart. (c) Height adjustment and sensing: lab jack on a telescoping mast, with a draw-wire sensor and digital readout. (d) Transmit patch and signal generator.}
\label{fig:platform}
\end{figure*}

\subsection{Measurement Setup and Procedure}
\label{sec:platform}

Field measurements were conducted in the Edgar Experimental Mine in Idaho Springs, Colorado, a former silver-and-gold mine operated by the Colorado School of Mines for education and research. The selected hard-rock drift is approximately 4.5~m wide and 3.3~m high (Fig.~\ref{fig:platform}(a)). Its rough, unlined rock surfaces, roof and wall wire mesh, gravel floor, and the presence of a parked mining vehicle and other equipment approximate practical deployment conditions. The mine Wi-Fi network was disabled during testing.

Patch~1 served as the transmitting antenna and Patch~2 as the receiving antenna. A Berkeley Nucleonics Model~845 microwave source supplied a 20~dBm continuous-wave signal at five frequencies: 2.40, 2.42, 2.44, 2.46, and 2.48~GHz. The transmitter, positioned beside the parked vehicle, was fixed at a height of 1.4~m (Fig.~\ref{fig:platform}(d)). The QuadBeam receiver was connected to a Tektronix RSA306B spectrum analyzer controlled by SignalVu-PC; a 200~MHz measurement span, 1~MHz resolution bandwidth, and ten-trace averaging were used. Reported received signal strength (RSS) values are powers at the analyzer input and are not corrected for cable loss or antenna gain.

A custom nonmetallic cart minimized nearby conductive structures (Fig.~\ref{fig:platform}(b)). It used a wooden frame, PVC handle, fiberglass axle, polymer bushings, and balloon tires. A telescoping mast provided coarse height adjustment, while a lab jack provided fine adjustment. A draw-wire displacement sensor reported height with 1~mm resolution (Fig.~\ref{fig:platform}(c)). All field height adjustments were performed manually.

Two measurement campaigns were performed. For the distance sweep, the transmitting and receiving antennas were maintained at heights of 1.4~m and 1.1~m, respectively, while the separation was increased from 8.3~m to 18.3~m in 0.1~m increments (101 positions). For the height sweeps, the cart remained stationary while the receiver height was adjusted over approximately 1.00~m to 1.20~m in increments of 5~mm or smaller. Measurements were repeated at a non-line-of-sight (NLOS) location 6~m from the transmitter, where a parked vehicle obstructed the direct path, and at line-of-sight (LOS) separations of approximately 15~m and 22~m. The 20~cm height range corresponds to about 1.6 wavelengths at the center of the test band. In total, 1{,}115 RSS readings were recorded across the measurement positions and frequencies.

\subsection{Results and Discussion}
\label{sec:results}

\begin{figure}[t]
\centering
\includegraphics[width=\columnwidth]{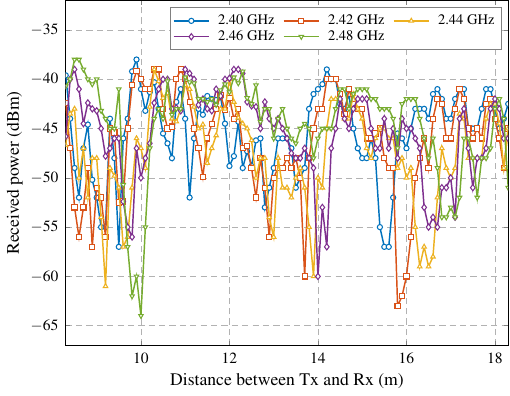}
\caption{Measured received power versus the separation between the transmitter and the receiver in the tunnel (transmitter power 20~dBm, transmit/receive antenna heights 1.4/1.1~m, tunnel cross section 4.5~m $\times$ 3.3~m).}
\label{fig:distance}
\end{figure}

\subsubsection{Spatial fading along the tunnel}
Fig.~\ref{fig:distance} shows pronounced position- and frequency-dependent fluctuations across the 10~m distance sweep. The fitted RSS-versus-distance slopes range from $-0.45$ to $+0.26$~dB/m, with no clear monotonic trend over the measured range. Carrier medians lie between approximately $-43$ and $-47$~dBm, with standard deviations of 4.3~dB to 5.2~dB. The peak-to-null RSS variation ranges from 19~dB to 26~dB (Table~\ref{tab:variation}), and a displacement of only 10~cm can change the received power by as much as 14~dB.

The fades are also frequency selective. At a separation of 10.0~m, for example, the received powers are approximately $-40$~dBm at 2.42~GHz and $-64$~dBm at 2.48~GHz. Deep-fade positions shift with carrier frequency. At 28 of the 101 sampled positions, at least one carrier is more than 10~dB below its own median. These observations are consistent with strong modal interference in the tunnel near zone~\cite{sun2010,zhou2015} and motivate local antenna repositioning after node deployment.

\begin{table}[t]
\caption{Peak-to-Null RSS Variation (dB) Measured During Each Sweep}
\label{tab:variation}
\centering
\setlength{\tabcolsep}{2pt}
\begin{tabular*}{\columnwidth}{@{\extracolsep{\fill}}lcccccc@{}}
\toprule
 & \multicolumn{5}{c}{Carrier (GHz)} & Lowest \\
\cmidrule{2-6}
Sweep & 2.40 & 2.42 & 2.44 & 2.46 & 2.48 & of five\textsuperscript{a} \\
\midrule
Distance, LOS & 19 & 24 & 22 & 21 & 26 & 21.5 \\
Height, NLOS, 6~m & 10 & 8 & 9 & 17 & 20 & 17 \\
Height, LOS, 15~m & 10 & 22 & 14 & 8 & 12 & 21 \\
Height, LOS, 22~m & 7 & 12 & 8 & 4 & 7 & 10 \\
\bottomrule
\end{tabular*}

\vspace{2pt}
\parbox{\columnwidth}{\footnotesize \textsuperscript{a}Range of the minimum RSS across all five carriers evaluated at each position; this need not equal the range of any one carrier.}
\end{table}

\begin{figure*}[t]
\centering
\subfloat[NLOS, horizontal distance 6~m.]{\includegraphics[width=0.32\textwidth]{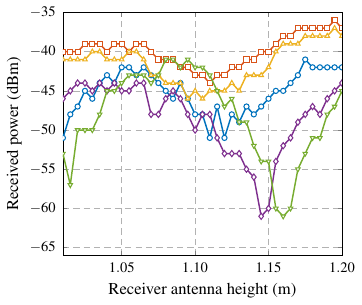}}\hfill
\subfloat[LOS, horizontal distance 15~m.]{\includegraphics[width=0.32\textwidth]{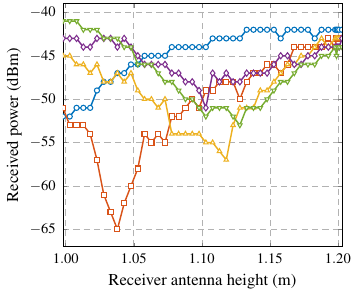}}\hfill
\subfloat[LOS, horizontal distance 22~m.]{\includegraphics[width=0.32\textwidth]{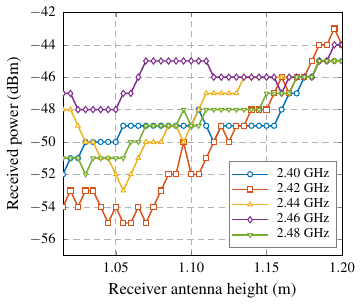}}
\caption{Measured received power versus receive antenna height at carrier frequencies of 2.40--2.48~GHz. The legend in (c) applies to all panels.}
\label{fig:height}
\end{figure*}

\subsubsection{Received-power recovery through height adjustment}
The height-sweep results in Fig.~\ref{fig:height} show that relatively small vertical displacements can substantially improve the received power. At the NLOS location, the 2.48~GHz RSS increased from $-61$~dBm at 1.160~m to $-41$~dBm at 1.095~m, corresponding to a 20~dB recovery over 6.5~cm. The 2.46~GHz carrier improved by 17~dB, while the three lower carriers varied by 8~dB to 10~dB. At the 15~m LOS position, height-dependent variations ranged from 8~dB to 22~dB; at 2.42~GHz, the received power increased from $-65$~dBm at 1.038~m to $-43$~dBm at 1.203~m. Variations at the 22~m LOS position were smaller, ranging from 4~dB to 12~dB.

Most small height increments produced modest changes: more than 90\% of the 5~mm adjustments changed the RSS by no more than 2~dB. This suggests that millimeter-scale positioning resolution is adequate for the tested conditions. Importantly, however, the best antenna height depends on both frequency and link geometry. At the NLOS site, the strongest height range for the 2.48~GHz carrier differs from that of the 2.40~GHz carrier; similar differences occur at the LOS sites. Hence, a single antenna position cannot be presumed optimal for every carrier or for independent links served by the same node.

Even when the objective is to maximize the worst received power among the five test carriers, height adjustment provides a meaningful benefit. Relative to the least favorable height, optimizing the minimum carrier power yields improvements of 17, 21, and 10~dB at the NLOS 6~m, LOS 15~m, and LOS 22~m sites, respectively (last column of Table~\ref{tab:variation}). Relative to the 1.10~m reference height of the distance sweep, the corresponding gains range from 6~dB to 10~dB. At the LOS sites, several carriers were strongest near the upper limit of the swept height range, suggesting that a longer travel range warrants investigation.

\subsubsection{Interaction among antennas}
To assess whether repositioning adjacent antennas affects an established link, the RSS was monitored while the three neighboring 2.4~GHz patches were moved and while the 900~MHz monopole was removed and reinstalled. Neither operation produced a noticeable change in the measured RSS. This qualitative field observation agrees with the low adjacent-patch coupling measured on the bench (Section~\ref{sec:testing}) and supports the feasibility of independent antenna positioning. It does not, however, constitute a full simultaneous multi-link throughput test.

\section{Conclusion and Future Work}
\label{sec:conclusion}

This study introduced QuadBeam, a compact antenna concept intended to improve the robustness of temporary underground mesh networks through independent antenna repositioning. Four 2.4~GHz patch antennas are mounted on the side faces of a shielded hub, with a 900~MHz monopole retained for backup voice communication. The preliminary prototype demonstrated adjacent-patch coupling below $-46$~dB and negligible observed disturbance of an active receiving link when neighboring antennas were repositioned.

Underground measurements in the Edgar Experimental Mine established substantial local spatial fading: across a 10~m drift segment, the five test carriers exhibited 19~dB to 26~dB peak-to-null RSS variations. Vertical repositioning over approximately 20~cm produced up to 20~dB recovery at an NLOS site and 22~dB at an LOS site. The height maximizing the RSS varied with frequency and measurement location, providing experimental motivation for independently optimizing antennas associated with different mesh links.

The results remain preliminary because only one mount is motor-driven and the mine campaign primarily evaluated received power for a single active link. Future work will equip all four antennas with closed-loop position control, increase the travel range, and investigate higher-gain printed Yagi elements. Longer-distance measurements, including mine crosscuts and intersections, will evaluate fading behavior outside the near zone. Finally, multi-node relaying and end-to-end throughput measurements will establish the practical network-level benefits of independent antenna positioning.

\section*{Acknowledgment}
The authors thank Jack Chiavacci, Harper Duong, Michael Farhy, Jessica Batson, Sean Artmann, Colin Thomas, and Jacob Moran of the Mines Capstone Design Program senior team for building the nonmetallic mobility cart used in the measurements.

\bibliographystyle{IEEEtran}
\bibliography{quadbeam_refs}

\end{document}